
\parindent=15pt
\parskip=25pt
\magnification=1200
\baselineskip=13pt

\rightline {KCL-TH-94-19}
\rightline{hep-th/9411029}
\rightline {October, 94}
\vskip 1in
{\centerline { {\bf{Physical States and String Symmetries  }}}}
\vskip 1in
{\centerline {\bf{Peter West}}}
\par
{\centerline {Mathematics Department}}
\par
{\centerline {King's College}}
\par
{\centerline {Strand}}
\par
{\centerline {London WC2R 2LS}}
\vskip 1in
{\underbar {Abstract}}
\par
It is shown that if the momenta belong to an integral lattice,
then every physical state of string theory leads
to a  symmetry of the scattering amplitudes. We discuss the role of
this symmetry when the momenta are those provided by the usual D.D.F
construction and show that the string compactified on the torus
associated with the self-dual Lorentzian lattice, $\Pi^{25,1}$ possess
the Fake Monster Lie algebra as a symmetry.
\vfil
\eject
\par
Even from the earliest attempt [1] to construct an interacting gauge
covariant string theory, it was clear that the string possesses
a symmetry that mixed mass levels and that the string three vertex
could be regarded as the structure constants of this symmetry. However,
even though we now possess consistent gauge covariant string theories
of the open [2] and closed [3] bosonic string, we lack a clear
understanding
of the symmetry principle that underlies string theory. One suggestion [20] was
that the vertices associated with the algebra of physical states based on the
unique self-dual Lorentzian Lattice
in 26 dimensions could play a role.
\par
In the group theoretic approach to string theory [5], the scattering amplitudes
are uniquely determined by a set of simple equations, called overlap equations.
These equations related how the operators in the theory of a given conformal
weight act on the string scattering vertices. The ovelap equations can be used
to derive equations, sometimes called integrated overlap equations, that relate
the contour integrals around Koba-Nielsen points of operators of  conformal
weight $d$ with suitable $1-d$ forms. For operators of conformal weight one,
the required forms are functions of which a special case is the constant. It
has been suggested [5] that such overlap relations may be regarded as
symmetries of the string theory. Indeed, the physical state tachyon vertex
operator was used [11] in this way to show that certain string vertices
satisfied the Hirota equation.  More recently [21],the conditions that the
conformal weight one operators, corresponding to the physical vertex operators,
imposed on the string vertex were considered when they were used in the above
mentioned equation . It was
shown [21] that, with some additional assumptions, they determined the tree
level tachyonic scattering amplitude.
\par
In this paper we further explore this  approach to symmetry in string
theory. We show, using the string three vertex,  that for any physical state
one can define a vertex
operator which, when integrated, can be used to construct a symmetry
of the string scattering amplitudes. For this result to hold,
however,  the action of the integrated vertex operator on the string
scattering vertex must be well defined. In general this is not the
case, but it holds if the momenta belong to an integral lattice.
There are at least two interesting situations where this is true;
one is when the string is compactified on a torus and the second is
when we work with with a string whose corresponding point particles
 have their momenta of the form $p-nk$ where $p\cdot p= 2,\
k\cdot k= 0,\ p\cdot k= 1$. This latter case arises in the well known
D.D.F construction of the physical states.
\par
The states of a quantum point particle belong to an irreducible
representation of the Poincar\'e group. Indeed, one frequently
turns this around and regards the quantum point particle as being
defined as an irreducible representation of the Poincar\'e group. The
usual bosonic string contains an infinite number of irreducible
representations of the Poincar\'e group. The string symmetry group can
be viewed as the symmetry group which organizes all these
representations into a single multiplet which comprises the string.
Since the early days of string theory it has been known that the
physical states  of the bosonic string are generated by the spectrum
generating algebra of the D.D.F operators [4]. The property of these
operators that made them so useful was that one could act with one of
them on any physical state and produce another physical state. In this
paper, we show that the D.D.F operators acting on the physical
states can also be interpreted as a symmetry of the scattering
amplitudes. As such, they lead to symmetries of the string theory
itself. Although the action of the D.D.F operators on the string
scattering vertices is well defined in the usual Lorentz frame used to
construct the corresponding states it is not well defined in any
Lorentz frame. Hence the symmetry only holds for particles with
momenta on the lattice associated with the D.D.F construction.
Possible ways of extending the construction to arbitary Lorentz
frames are discussed.
\par
Another situation in which one finds the above symmetry is valid is on
string compactifications on a torus. A particularly interesting case
is the completed compactification of the string on the torus
corresponding to the unique self-dual Lorentzian lattice $\Pi
^{25,1}$. In this case the string theory possess the Fake Monster
algebra as a symmetry group and the $N$ string scattering vertex
and the $N$ string S-matrix itself can be viewed
as an
invariant tensor and a kind of Casimir of this group respectively.
\par
The calculations in this paper are carried out in the group theoretic
approach [5 ]. The principal object in this approach is the scattering
vertex. For  tree level scattering,  the scattering vertex
$V^{(N)}$ is related to the
 scattering of N strings
$\vert\chi >_j, \ \ j=1,\dots, N$ by
$$
W^{(N)} = \int {\prod\limits_i}^\prime d z_i V^{(N)}
\prod
\limits^N_{j=1} \vert\chi >_j.
\eqno(1)$$
where  $z_i$ are the
Koba-Nielsen coordinates.  The $N$ string vertex is
defined to satisfy the ``overlap equations''
$$
V^{(N)} R^{(j )}(\xi^j) (d \xi^j)^d = V^{(N)} R^{(i)} (\xi^i) (d
\xi^i)^d
\eqno(2)$$
for any operator of conformal weight $d$.  These imply
the
integrated ``overlap equations''
$$
V^{(N)} \sum\limits^N_{j=1} \oint d \xi^j R^{(j)} (\xi^j)
\phi = 0
\eqno(3)$$
where $\phi$ is a $1 - d$ form.  Strictly, one should define the
vertex using equation (2) only for the basic  operators of the
theory, however it has been found that one can then show equation (2)
then follows for all the conformal operators of the theory usually
considered.  The above equation also holds for loop scattering
vertices, but in this case, the vertex also obeys additional overlap
equations, one for each loop.
\par
The physical states of the bosonic string satisfy
$$
Q \vert \varphi > = 0\ \ , \ \ b_0 \vert\varphi > = 0
\eqno(4)$$
The non-trivial cohomology classes belong to the states [6]
$$
\psi c \vert 0>\ \ ,\ \ \vert 0 >
\eqno(5)$$
where $\vert 0 >$ is the $SL(2,{\bf{R}}) $ invariant vacuum
and $\psi$ is a functional of $x^\mu$ alone which is
subject to $L^x_n \psi = 0\ \ n \ge 1\ \ ;\ \ (L^x_0 - 1)
\psi = 0$.
\par
Given any physical state $\vert \varphi >$ we can use the
three string vertex $V^{(3)}$, with ghost included [7],
 to form a
corresponding operator; we act with $\vert \varphi >$ on leg three to
form
$$
V_\varphi (z) \equiv V^{(3)} \Omega^{(2)} z^{L_0^{(1)}}
z^{-L_0^{(2)}}( b_0^{(1)} - b_0^{(2)}) \vert \varphi > _3
\eqno(6)$$
where $\Omega^{(2)}$ is the twist operator applied on leg two, whose
precise form [8] depends on the choice of local coordinates
$\xi^j$ used to define the vertex.  For the old Canneschi,
Schwimmer, Veneziano vertex [9], $\Omega = e^{-L_1}
e^{i\pi L_O} $.  Since the B.R.S.T. charge is the
integral
of a weight one conformal current, the N vertex satisfies
$$
V^N \sum\limits^N_{j=1} Q^{(j)} = 0.
\eqno(7)$$
Using the relation $\{ b_n, \psi \} = L_n$, one finds
that the above equation implies that
$$
V^{(2)}_\varphi (z) (Q^{(1)} + Q^{(2)}) = z {d\over dz}
V^{(2)}_\varphi (z)
\eqno(8)$$
Clearly, the vertex $V_\varphi^{(2)}$ also satisfies
$V_\varphi^{(2)} (z) (b_0^{(1)} -b_0^{(2)}) = 0.$
{}From this two vertex we can define an operator
$V_\varphi(z)$ by turning around line 1 and then
identifying the 1 and 2 lines:-
$$
< \phi \vert V_\varphi \vert \chi > =
V^{(2)}_\varphi \vert \phi>_1 \vert \chi >_2
\eqno(9)$$
for all $\vert \phi >$ and $\vert \chi >$.  The
relations above then become
$$
[V_\varphi(z), Q] = z {d\over dz} V_\varphi (z)
\eqno(10)$$
$$
[b_0,\ V_\varphi (z) ] = 0
\eqno(11)$$
Consequently, the operator
$$
W_\varphi = \oint {dz\over z} V_\varphi(z)
\eqno(12)$$
commutes with $Q$ and $b_0$. It follows that  $W_\varphi$ acting
on a physical states takes it to another physical state. If the
physical state $\varphi$ is of the form $\vert \varphi > = Q\vert
\wedge >$ then
$$
W_\varphi = [Q, W_\wedge]
\eqno(13)$$
A consequence of the fact that the physical state $\vert
\varphi >$ is annihilated by $L_n ; \ \ n\ge 0$ is
that $V_\varphi{(z)}$ an operator of conformal weight
one.  To demonstrate this result one considers the
integrated overlap of equation (3) taking $R=T$ and $\phi$ to be a
vector field that is analytic everywhere except  at the
Koba-Nielsen points associated with legs one and two, where it may
have poles, and it also vanishes at the Koba-Nielson point
corresponding to string three.
\par
One could derive analogous results for an operator
$V^\prime_\varphi$, obtained by the above procedure with the
exception that one did not insert in equation (6) the $(b_0^{(1)} -
b_0^{(2)})$ factor.  This operator, $V^\prime_\varphi$
will commute with $Q$ and be of weight zero.
These results are consistent with the association
between physical states and the vertex operator used in the conformal
field theory approach to scattering in string theory.
\par
The above analysis used a formalism which included ghosts in all the
 vertices and Virasoro generators. We can also
work without ghosts, that is in the old covariant formalism.
Following almost identical arguments one finds a $V_\varphi$
that is an operator of conformal weight one. It's integral
$W_\varphi$ takes physical states to physical states. For the rest
of this paper we will, for convenience, work in the formalism without
ghosts since in this case the scattering amplitudes, as shown in
equation (1),  do not require ghost insertions which complicate the
form of the overlaps.
\par
It now follows
from the overlap of equation (3) that for any physical state
$\vert
\varphi >$, the N string vertex obeys the equation
$$
V^{(N)} \biggl\{\sum\limits^N_{j=1} W_\varphi^{(j)}
\biggr\} = 0
\eqno(14)$$
What is not immediately apparent however, is that this equation
only holds if the contour integrals contained in $W_\varphi$ are
well defined.  This will be the case if the action of
$V_\varphi{(z)}$ on $V^{(N)}$ induces only poles in $z-z_j$ instead
of more singular behavior.
\par
Examining the form of N string scattering of equation (1) one
finds that it is invariant under
$$
\delta \vert \chi > = W_\varphi \vert \chi >
\eqno(15)$$
provided the action of $W_\varphi$ on $\vert \chi >$ is
also well defined.    Consequently, it is a symmetry of
both the states and the scattering amplitudes and so is a symmetry
of string theory.
\par
The most usual variation of a field found in quantum
field theory is the product of the field itself, a representation of the
generator of the group and the parameter.
The analogue with the variation found above is to interpret
the physical state as the parameter and the three vertex
as the group representation of the generator. Since the physical
states lead to the generators of the group and we act with these
 on physical states, we are using an adjoint action of the
group and so we should regard the three vertex as the structure
constants.
\par
It is instructive to derive equation (4) for the photon states.
 It is straight-forward
to follow  the procedure leading to equation (12) taking
$\varphi $ to be the photon state. One finds that the corresponding
$W_\varphi$'s are  the  D.D.F  operators [4],
$$
A^i_n (k) = \oint \ dz\ e^{ink\cdot Q{(z)}}{\bf  P}^i{(z)}.
\eqno(16)$$
where ${\bf P}^i = \partial Q^i,\ \  k^2 = 0$ and we have chosen
to work in the Lorentz frame where $k^i = 0$.
The N string vertex obeys the overlap equation
$$
V^{(N)} Q^{(i)\mu} (\xi^i) = V^{(N)}
Q^{(j)\mu}(\xi^j)\eqno (17)
$$
Differentiating with respect to $\xi^i$ we find that
$$
V^{(N)} {\bf P}^{(i)\mu}(\xi^i) = V^{(N)} {\bf P}^{(j)\mu} (\xi^j) {d
\xi^j\over d\xi^i}.
\eqno(18)$$
Equation (17) implies that
$$\eqalign{
& V^{(N)} Q^{(i)\mu_1} (\xi_1) Q^{(i)\mu_2} (\xi_2)\dots
Q^{(i)\mu_n} (\xi_n)\cr
& = V^{(N)} Q^{(j)\mu_n} (\xi_n) \dots Q^{(j)\mu_1} (\xi_1).}
\eqno(19)$$
If we consider $k_\mu Q^\mu$ for $k^2 = 0$, we have no
normal ordering problems and
$$
V^{(N)} [ik\cdot Q^{(i)}(\xi^i)]^m = V^{(N)} [ik \cdot
Q^{(j)} (\xi^j)]^m
\eqno(20)$$
Dividing by $m!$ and summing we find that
$$
V^{(N)} \exp \bigl[ik\cdot Q^{(i)} (\xi^i)\bigr]= V^{(N)}
\exp\bigl[i k \cdot Q^{(j)} (\xi^j)\bigr]
\eqno(21)$$
Applying equation (18) and taking $k \to nk$ we find that
the photon emission operator obeys the equation
$$
V^{(N)} e^{ink \cdot Q^{(i)}(\xi^i)} {\bf P}^{(i)k}(\xi^i) =
V^{(N)} e^{ink\cdot Q^j(\xi^j)}{\bf P}^{(j)k} (\xi^j) {d
\xi^j\over d\xi^i}
\eqno(22)$$
Integrating this overlap equation we find that
$$
V^{(N)} \sum\limits^N_{j=1} A^{(j)i}_n = 0.
\eqno(23)$$
Since the D.D.F operators
 are the contour integral of the physical photon
emission operator, which is an  operator of conformal weight one, they
 commute with the
$L_n$. Consequently, the transformation
$\delta \vert \chi> = A_n^k \vert \chi >$ takes us from one physical
state to another, and equation (23) tells us that this is also a
symmetry of the string scattering amplitudes.
\par
One can also establish [11] a related result to equation (21) for any
momentum $k^\mu$, whereupon one finds an additional factor of
${({d \xi^j\over d\xi^i})}^{k^2\over 2}$. Consequently, for $k^2=2$,
ie tachyons, one finds an equation similar to equation (23), except
with no
${\bf P}^k$, and so an analogous symmetry.
\par
It is well known, at least for D=26, that all the physical states with
positive norm are of the form
$$
\prod\limits_{\{\lambda_{n,i}\}} (A^i_{-n,i})^{\lambda_{n,i}}
\ \ \vert 0, {p^\prime} >
\eqno(24)$$
where ${( {p}^\prime)}^2 = 2,\ \
{p} ^\prime. {k} = 1.$  Since, the D.D.F states generate all
states it follows that equation (23) holds for all physical states.
Although one can work with D.D.F operators  alone, it is
perhaps more natural to work with all the physical states as
we first did.
\par
To clarify the nature of the above symmetry, it is
instructive to recall the analogous steps for the point
particle.  The quantum point particle transforms as an
irreducible representation of the Poincar\'e group.
The unitary irreducible representations of the
Poincar\'e group are found using the method of induced
representations which we now summarize for the massless
case.  Given a particle of momentum $k^\mu, \ k^2 = 0$ we can
choose its components to be $k^{-} = {1\over \sqrt 2.}
(k^{D-1} - k^0)=1, k^+ = k^i = 0$.  This choice is
preserved by the ``little group'' $ISO (D-2)$ generated
by the momentum operators $P^\mu , \ J^{ij}\ ;\ i,j =1,
\dots, D-2$ and $J^{i+}$.  We now choose an irreducible
unitary representation of this group.  Since the $J^{i+}$
generate a non-compact group, any unitary representation,
$\varphi$ must be trivial under these generators; $
J^{+i} \varphi = 0.$  One is then left with $S0(D-
2)$.  The representation of the full Poincar\'e group is
then found by boosting the $\varphi$ according to the
usual induced procedure.  We can,  for example, choose the
fundamental representation, $\varphi^i, \ i=1, \dots, D-2$
which corresponds to the photon.
\par
Physicists usually find it useful not to work with the
above representation, but to extend it by embedding it into a
larger representation.  For the photon,  we
consider $\varphi^\mu; \ \mu = 0, 1, \dots, D-1$ which
transforms in the fundamental representation of $SO(D-1,
1)$.  For massless particles, this embedding can
be performed in two steps. In the first step, we consider the  $SO(D-2)$
invariant subspace given by
$\varphi ^{\hat \mu } = \{ \varphi ^i, \varphi ^- \}$. In this space we
 release the unitarity restriction mentioned above,
but replace it by  the equivalence
relation
$\varphi^{\hat \mu} \sim \varphi^{\hat \mu \prime}$ if
$\varphi^{\hat \mu}= \varphi^{\hat\mu \prime}+ k^{\hat \mu}
\Lambda$.  This is just the gauge transformation which in
effect removes $\varphi^-$.
Finally, we can consider the full space $\varphi^\mu$, but
impose the condition $k\cdot \varphi = 0$ which restricts
one to the formulation just discussed.
\par
Let us now consider the analogous steps for string
theory.  It is well known [10] that the D.D.F operators
admit a Lorentz extension
$$
A^\mu_n(k) = \oint: e^{ink.Q(z)}{\bf P}^\mu (z) : dz
-{n\over 2} k^\mu F_n (k)
\eqno(25)$$
where
$$
F_n(k) = \oint dz {k\cdot : \partial {\bf P}\over k\cdot {\bf P}}
e^{ik\cdot Q}: \eqno (26 )
$$
These operators obey the algebra
$$\eqalign{
 [A^\mu_n(k), A^\nu_m (k)] = &n\ \ \delta_{n+m,0} P\cdot k
\eta^{\mu\nu} + m \ k^\mu A^\nu_{n+m} (k)\cr
&-n k^\nu A^\mu_{n+m}(k) + n^3 k^\mu k^\nu \delta_{n+m,0}
P\cdot k \cr}  \eqno(27)
$$
where we have used the relation $k\cdot A_n(k) =
\delta_{n,0} P\cdot k$.
\par
One choice of little group is  that generated by $P^\mu,
A^i_n(k)$ and $J^{ij},\ \ ,i,j=1, \dots, D-2$ with $k^- =1,\
k^i = k^+ = 0$.  Acting on the momentum state $\vert
p^\mu>$ with momentum $p^+ = p^- = 1\ ,\ p^i=0$ with the $A^i_n$
we generate all the physical states of the string with
positive definite norm.  We note that we can not include
more of the Lorentz group since these will not leave both the
massless vector $k^\mu$ and the massive vector $p^\mu$
inert.
\par
We could also extend the choice of little group to be
$P^\mu, A^i_n, A^-_n$ and $J^{ij}$.  The generators $A^-
_n$ obey the relations implied by equation (27), but it is
preferable to work with
$$
R_n = -A^-_n - {1\over 2} \sum \limits^{D-2}_{i=1}
\sum\limits^\infty_{p=-a} : A_{n-p}^i  A^i_n: -{(D-
2)\over24} \delta_{n,0}
\eqno(28)$$
since it obeys the simpler algebra
$$\eqalign {
[R_n, A^i_m] &= 0\cr
[R_n, R_m] &= (n-m) R_{n+m} + {n^3\over12} (26-D)
\delta_{n+m,0}}
\eqno(29)$$
The $R_n$ obey a Virasoro algebra with central charge $26-
D$ which vanishes in $D = 26$, a choice we now adopt.  In
a unitary representation, $R_n$ must therefore be
trivially realized and so one is left with the previous
D.D.F. states.
As for massless representations of the Poincar\'e group,
we can extend the represention to include  states which
have non-trivial $R_{-n}$ factors and consider two
states as equivalent if they differ by a state that
includes any $R_{-n}$'s acting.
\par
To carry out the analogue of the final step, that is  to obtain a
manifestly Lorentz covariant formulation, we add to our
operators, an operator $\phi_n(k)\ ,\
k^2=0$  which satisfies
$$\eqalign{
[\phi_n(k)\ ,\ A^\mu_m(k)] &= n k^\mu \phi_{n+m}\cr
[\phi_n(k)\ ,\ \phi_m(k)] &= 0}
\eqno(30)$$
A representation of $\phi_n$ is given by [10]
$$
\phi_n = \oint {dz\over z} e^{ink\cdot Q(z)},
\eqno(31)$$
 however, we can regard $\phi_n$ as an abstract
operator at this point.
\par
We now consider states in an enlarge configuration space
of the form
$$
\vert\{K,\lambda\}. = \prod\limits_{\{K_m\}} (\phi_
{-m}{(k)})^{K_m}
\prod\limits_{\{\lambda_p\}} (R_{-p}{(k)})^{\lambda_p} \prod \limits_
{\{\lambda_{n,i}\}} ( A^i_{-n} (k) )^{\lambda_{ n,i}} \ \ \vert 0,
p>.
\eqno(32)$$
where
$$
\phi_n \vert 0, p> = 0 = R_n \vert 0, p>  = A^i_n\ \ \vert 0,p>,
n\ge 1
\eqno(33)$$
and $k\cdot p=1$.  Clearly, this enlarged representation
contains the representation of equation (24) by simply
omitting the $\phi_{-m}$ and $R_{-p}$ factors.
\par
 In fact, all the states
of equation (33) are linearly independent.  The simplest
way to demonstrate this fact is to note that the determinant
of the matrix
$$
M(\{K,\lambda\}, \{K^\prime, \lambda^\prime\}) =
<K,\lambda | K^\prime, \lambda^\prime>
\eqno(34)$$
is non-zero.  The proof is similar to that given by Thorn
[13], albeit for a different set of operators.  Essentially,
one can adopt an ordering of the oscillator factors such
that $M$ has non-zero entries on the diagonal from the
lower left to the upper right and vanishing entries below
this diagonal.  This occurrence stems from the fact that
$$
<0,p\vert \phi_{n_1}\ \phi_{n_2} \ \phi_{n_3}\dots
\phi_{n_s} \vert 0,p> \eqno (35)
$$
vanishes unless $n_1 = n_2= n_3= \dots = n_s =0$.
\par
We can also consider the linearly independent states of the form
$$
\prod \limits_{\{\tau\}} (\alpha^\mu_{-
n})^{\tau_{n,\mu}} \vert 0,p>\eqno(36)
$$
where $\alpha^\mu_n \vert 0,p> = 0, n\ge 1.$  The number
of states at each level is determined only by the number
of oscillator operating on the vacuum.  There are
therefore the same number of states in the Hilbert space whose
states are the form given in  equation (36) as
in the Hilbert space whose states are of the form of those of
 equation (32).  Indeed, using the
explicit formulas for $A^\mu_n$ and $\phi_n$ in terms of
$\alpha^\mu_n$ we may construct the latter Hilbert space
from the former.
\par
To recover the original representation in this larger
Hilbert space we must find the projection conditions.
The long history of string theory has provided us with
these objects.  To project onto the states of equation (24)
we can use the Brink-Olive projection operator [14],
while if we wish to restrict only to the states of
equation (32) with no $\phi_n$'s we use the Virasoro conditions
$$
(L _n -\delta _{n,0}) \vert \{K,\lambda\} > = 0 \ \ n\ge 0
\eqno (37)
$$
where $L_n= {1\over2} \sum: \alpha^\mu_{n-p}
\alpha^\mu_p:$
\par
To demonstrate this we define the operators
$$
\phi_{m,n} = \oint {dz\over z} z^m e^{ink\cdot Q}
\eqno(38)$$
which have the following commutator with $L_n$;
$$
[L_n, \phi_{m,p}] = -(n+m) \phi_{n+m,p}
\eqno(39)$$
Acting with
$$
(L_1)^{K_1} \dots (L_m)^{K_m} \eqno (40)
$$
on $\vert \{K, \lambda \}>$ we can reduce all the $\phi
_n$'s to $\phi_{r,n}$'s  which either annihilate due to the relation
$$
\phi _{m,n} \vert 0,p> = 0 \eqno (41)
$$
unless $m+n\le 1 $ or, if $r+n =0$, give a factor one as a result
of the relation
$\phi_{-n,n}
\vert 0, p> =\vert 0, p>$ . Consequently, one is left with only the
$ A^\mu_{-n}$ oscillators acting on $\vert 0, p >$.  However,
according to equation (37) this should vanish which can only happen
if there were no $\phi_{-n}$ factors in the original
state.  Hence, the Virasoro conditions do achieve the
relevant projection conditions.
\par
Thus we see there is a close parallel between the theory of
induced representations of the Poincar\' e group for the point
particle and the structure of the states on string theory. In the
string theory case  we have extended the little group
algebra to include the
$\phi _n$'s.  This allows us to construct the $L_n$'s which commute
with the original algebra.  The
$L_n$ constraints are are usually derived from the Nambu action after
first quantization, however, the above derivation provides a more algebraic
derivation which it would be desirable to make more systematic.
\par
The action of $V_\varphi$ on the $N$ string
vertex is not in general well defined. If the string carries a
momentum $p_j$ on leg j and the physical state carries momentum $k_j$
, $V_\varphi$ will generate a term of the form ${(\zeta^j
-z^j)}^{p_j\cdot k_j}$  when it acts on the scattering vertex.
This term when integrated will
 only be well defined if
$p_j\cdot k_j\in {\bf Z}$. If , however, all the states have momenta
of the form
$p-nk,\ n\in {\bf Z} $ then the action is well defined and we have a
good symmetry. In the Lorentz frame in which the D.D.F
operators are usually constructed, the momenta of the states do
have the above form and so the above symmetry generated by
physical states is a symmetry of the string theory in this Lorentz
frame. In a general Lorentz frame the action of the symmetry
generators is not well-defined since $p_j\cdot k_j$ is not an
integer. It is perhaps not unreasonable to try to build a string theory
in which this symmetry generalizes to all Lorentz frames. There are
several approaches to this problem. One could drop the requirement of
possessing commutation relations. An example of an algebra whose
currents do have operator product expansions that involve only
poles is that generated by parafermions [12] , although even in this
case the singularities in the operator product expansions are of the
form
$(z-w)^{p +{ q\over N}}$ where  $p,q,N \in {\bf Z}$.  An alternative
is to try to extend the theory so as to still retain a Lie
algebra, one such possibility is to added twisted fields [16] to the
theory. It has been found [17] that  one can  can construct
projective invariant
 off-shell vertex operators, in 26
dimensions, using these fields and one may
hope that these vertex operators have operator product expansions that
can be used to define Lie algebras. It is worth recalling that these
fields were originally introduced in order to construct off-shell string
scattering. One could also attempt to recover the known gauge
covariant string theories from this starting point.
\par
The symmetry will also be valid when one compactifies left and right
sectors independently on a torus, since the associated momenta belong
to $\Lambda \cap \Lambda ^\star$ where  $\Lambda$ and $\Lambda ^\star$
are the associated lattice and its dual. A particularly,
interesting case is the complete compactification of the closed
bosonic string on the self-dual Lorentzian lattice $\Pi^{25,1}$. The
algebra of physical states associated with this lattice has
been studied in reference [15] and is referred to as the Fake Monster
Lie algebra.  This algebra is  generated not only by a subset of the states
corresponding to roots of length two, but also by those corresponding
to integer multiples of the null vector
$w=(0,1,2\ldots,24;70)$. One could construct the physical states
using a D.D.F construction as was done for low levels for the
algebra $E_{10}$ [18]. Since all the momenta belong to  $\Pi^{25,1}$
the action of the integrated  vertex operator of equation (12) on the
$N$ string vertex is always well defined and as a result every physical
states generates  a vertex operator that is a symmetry of the
scattering amplitudes.  Hence, we conclude that the bosonic string
compactified on the torus associated with the self-dual Lorentzian
lattice
$\Pi^{25,1}$ has the Fake Monster algebra as a symmetry. Indeed an $N$
string vertex can be viewed as an invariant tensor and the $N$ string
S-matrix as a kind of Casimir of this algebra.  Given that the vertices
are  uniquely determined by the $Q^\mu$ overlap equations it is
perhaps not unreasonable to suggest
that one could determine the properties of this string
solely from demanding that the Fake Monster algebra be a symmetry. Some
progress in this direction has been made in reference [21] and it
would be interesting to investigate this suggestion further  and
perhaps use it to try to determine any non-perturbative behavior the
string possess.
\par
One might regard the symmetry discussed in this paper as part of the
deeper symmetry that string theory is thought to possess. A useful
analogy can be made with general relativity where one might suppose
that prior to the discovery of general relativity physicists had a
method of finding solutions to Einstein's equations without knowing
what these equations were or understanding general coordinate
invariance. One would find that certain special solutions possessed
more symmetry than others, from our modern perspective these would
result from the presence of Killing vectors, but from the perspective
 of the early physicists these additional symmetries might provide
 clues to the principle of general coordinate invariance [19]. The same
might be true of the of the string compactified on the $\Pi^{25,1}$
lattice in that it could be used as a toy model to gain insights into
string theory.
\par
It would also be of interest to consider the $\alpha \to 0$ limit of the above
algebra. Taking this limit in the most naive manner implies that the D.D.F
operators commute and as a consequence all the physical states except for the
tachyon become null. While this calculation is very heuristic it does lead to
the conclusion that the string at high energy has fewer degrees of freedom, in
agreement with a number of other considerations.
\parskip=5pt
\medskip
{\bf Acknowledgement}
\par
I wish to thank Reinhold Gebert, Paul Howe and Hermann Nicolai
for  discussions.
\par
{\bf References}
\item{[1]} A. Neveu and P. West Nucl Phys B268 (1986) 125.
\item{[2]}  A. Neveu and P. West Nucl Phys B293 (1986) 266,
\hfil\break
   E. Witten  Nucl Phys B268 (1986) 253.
\item{[3]}  A. Neveu and P. West Nucl Phys B293 (1986) 266,
\hfil\break
  B. Zwiebach "Closed string field theory: an introduction", preprint
Hep-th/9305026.
\item{[4]}	E .Del Giudie, P. Di Vecchia and S. Fubini; Ann. Phys. 70
(1972) 378.
\item{[5]} For a review of the group theoretic approach see
P.West, Nucl. Phy. B (Proc. Suppl) 5B (1988) 217, Nucl. Phys. B320
(1989) 103.
\item{[6]}	M. Kato and K. Ogawa; Nucl. Phys. B212 (1983) 443,
\hfil\break
 A. Neveu, H. Nicolai and P. West; Phys. Lett. 175B (1986)
307, \hfil\break
 M. D. Freeman and D. Olive; Phys. Lett. 175B (1986)
151.
\item{[7]} A. Neveu and P. West; Nucl Phys B278 (1986) 601.
\item{[8]} A. Neveu and P. West; Commun. Math. 119 (1988) 585.
\item{[9]} L. Caneschi, A Schwimmer and A. Veneziano,
Phys. Lett, 30B (1969) 351.
\item{[10]}	R .C .Brower; Phys. Rev. D6 (1972) 1655.
\item{[11]}	B. Nilsson and P. West, Commun. Math. Phys. 145 (1992)
329.
\item{[12]} A. Zamolodchikov Theot. Math. Phys. 65 (1989) 1205.
\item{[13]}	C. B .Thorn; Nucl. Phys. B248 (1984) 551.
\item{[14]} L. Brink and D. Olive; Nucl. Phys. B56 (1973) 253.
\item{[15]} R. Borcherds Advances in Mathematics 83 (1990) 30.\hfil\break
For a review see
R. Gebert, Int. Jour. Mod. Phys. A8 (1993) 5441.
\item{[16]} E. Corrigan and D. Farlie, Nucl. Phys. B91 (1975) 397.
\item{[17]} M. Bershadski; Int. Jour. Mod. Phy. A1 (1986) 443.
\item{[18]} R. Gebert and H. Nicolai "On $E_{10}$ and the D.D.F
construction" preprint Desy 94-106.
\item{[19]} This analogy was put to me by Hermann Nicolai.
\item{[20]} A. Neveu and P. West; Phys. Lett. 179B (1986) 235.
\item{[21]} G. Moore, "Symmetries of the Bosonic String S-matrix"
hep-th/9310026 and hep-th/9404025.

 \end